\documentclass[10pt,journal,compsoc]{IEEEtran}
\usepackage{amsmath,amsfonts}
\usepackage{array}
\usepackage{textcomp}
\usepackage{stfloats}
\usepackage{url}
\usepackage{verbatim}
\usepackage{graphicx}
\usepackage{subfigure}

\usepackage{booktabs} 
\usepackage{wrapfig}
\usepackage[utf8]{inputenc}
\usepackage[left=0.7in,
            right=0.7in,
            top=0.7in,
            bottom=0.9in,
            footskip=.25in]{geometry}

\usepackage{pdfpages}
\usepackage{todonotes}
\usepackage{algorithm}
\usepackage{algpseudocode}
\usepackage{textcomp}
\usepackage{graphicx,subcaption}
\usepackage{tikz}
\usepackage{xspace}
\usepackage{cite}
\usepackage{csquotes}
\usepackage{enumitem}
\usepackage{tabularx}
\usepackage{hyperref}
\usepackage[capitalize]{cleveref}
\usepackage{multirow}
\usepackage{balance}

\hypersetup{
  colorlinks   = true, 
  urlcolor     = black, 
  linkcolor    = black, 
  citecolor   = black 
}

\newcommand{\sol}{{ReStorEdge}\xspace}

\errorcontextlines\maxdimen

\crefname{algocf}{alg.}{algs.}
\Crefname{algocf}{Alg.}{Algs.}

\def\BibTeX{{\rm B\kern-.05em{\sc i\kern-.025em b}\kern-.08em
    T\kern-.1667em\lower.7ex\hbox{E}\kern-.125emX}}
\begin{document}
\title{ReStorEdge: An edge computing system with reuse semantics}
\author{
\IEEEauthorblockN{Adrian-Cristian Nicolaescu}
\IEEEauthorblockA{University College London
a.nicolaescu@ee.ucl.ac.uk}\\
\and
\IEEEauthorblockN{Spyridon Mastorakis}
\IEEEauthorblockA{University of Notre Dame 
mastorakis@nd.edu}\\
\and
\IEEEauthorblockN{Md Washik Al Azad}
\IEEEauthorblockA{University of Notre Dame
malazad@nd.edu}\\
\and
\IEEEauthorblockN{David Griffin}
\IEEEauthorblockA{University College London
d.griffin@ucl.ac.uk}\\
\and
\IEEEauthorblockN{Miguel Rio}
\IEEEauthorblockA{University College London
miguel.rio@ucl.ac.uk}
}

\markboth{Transactions on Emerging Topics in Computing}{Adrian-Cristian Nicolaescu, Spyridon Mastorakis, \MakeLowercase{\textit{(et al.)}}: ReStorEdge: An edge computing system with reuse semantics}

\maketitle

\begin{abstract}


This paper investigates an edge computing system where requests are processed by a set of replicated edge servers. We investigate a class of applications where 
similar queries produce identical results. To reduce processing overhead on the edge servers we store the results of previous computations and return them when new queries are sufficiently similar to earlier ones that produced the results, avoiding the necessity of processing every new query. 
We implement 
a similarity-based data classification system, 
which we evaluate based on real-world datasets of images and 
voice queries. 
We evaluate a range of orchestration strategies to distribute queries and cached results between edge nodes 
and show that 
the throughput of queries over a system of distributed edge nodes can be increased by 
25-33\%, increasing its capacity for higher workloads.


\end{abstract}


\section{Introduction}

\IEEEPARstart{T}{oday's Internet} is generating and using increasing quantities of data at the edge. 
This is due to a variety of mobile, mixed reality, Internet of Things (IoT) and smart city applications generating large amounts of requests, for in-network computation~\cite{shi2016edge}. Edge computing solutions have been proposed and have started being deployed in real-world scenarios to address this change~\cite{hung2018videoedge, 9488804}. Because of the massive amounts of requests and data travelling every second through the Internet, many edge applications are bound to operate on similar or correlated data, which is a feature that can be exploited for serving multiple queries with the same results~\cite{guo2018potluck, al2022promise}.
In this context, the field of \emph{computation reuse} is emerging, 
which proposes the sharing of processing results among applications and services in application cases in which similar input data can be considered reusable, while used in different instances of the same application, or, more generally, even in other applications/services within the same context (as long as a pre-assessment is done and the application providers are clear and transparent with these terms, with respect to their users.

Some of the types of generally-well-correlated data are videos, same-subject, or same-device photos and same-user/same-household smart assistant commands. There are others, like vehicle-based measurements, certain building control and security systems measurements, or even application-based service requests, but we motivate this work with the former use cases. All of these contribute to the collections of edge-generated and edge-used data, that could also be computed and potentially reused within the edge, for increasing throughput, improving latency, and overall improving resource availability, round-trip service latency and data distribution.

Mechanisms for reducing the computational overhead and latency of similarity searches have been widely studied in the literature~\cite{FALCONN,10.14778/2556549.2556574,10.5555/1325851.1325958,10.1145/2020408.2020578,10.1145/1873951.1874168,9762397} and are not investigated further in this study.

The overall problem statement we are addressing in this paper can be formulated as follows: Given a set of distributed edge nodes that are able to process user queries for a given application, queries should be directed to edge nodes so that: 1) the opportunity for reuse of prior cached results for similar queries is maximised; 2) load is balanced across edge nodes, given the processing implications of the resulting mix of reused and fully processed queries, to minimise query response time and maximise query throughput, also optimising resource use in the process.

In this paper we propose an edge framework, called \sol, which improves in-network processing to work towards session-less data processing. 
We make the following contributions:

\begin{itemize}[leftmargin=0cm,itemindent=0.3cm,labelwidth=\itemindent,labelsep=0cm,align=left, noitemsep, topsep=0cm]

\item We propose the use of an architecture for processing queries on a set of edge-based, storage- , computing- and network-enabled servers, called Edge Data Repositories (EDRs);



\item We propose a set of orchestration strategies to manage the distribution of queries between EDRs to minimise the computational load on EDRs, maximise the reuse of previous results and maximise Quality of Experience (QoE) for user queries by minimising query latency;

\item We evaluate the proposed architecture by, firstly, implementing a prototype of \sol to quantify the benefits of computation reuse in a range of applications; and, secondly, using the results of the prototype evaluation in a wider simulation of distributed EDRs to evaluate system performance under different orchestration strategies.

\end{itemize}

The rest of our paper is organised as follows. In Section~\ref{sec:related}, we present a brief background on storage and computation frameworks, discuss prior related work, and present relevant use cases. In Section~\ref{sec:Overview}, we present our \sol system design. In Section~\ref{sec:eval}, we present our experimental evaluation,
and finally, in Section~\ref{sec:conclusion}, we discuss our future work and conclude our work.

\section{Prior Related Work and Use Cases}
\label{sec:related}

\subsection{Edge-based Data Storage and Computation Offloading}


In \cite{9488804, 10.1145/3359993.3366644} and \cite{9762397}, the authors provide insights into Edge networked storage systems and how these could be managed, towards timely function execution. However, the proposed management strategies consider how data storage could be managed, and not how both processing functions and data storage resources could be orchestrated, in order to reduce both resource usage and end-to-end latency. At the same time, prior work has explored different strategies for computation offloading from user devices to the edge of the network~\cite{10.1145/3434770.3459740, 10.1145/3318216.3363303, Krol2018}. Together, Edge-based data storage and computation offloading~\cite{10.1145/3359993.3366644} can form the basis for frameworks that facilitate the operation of applications at the edge of the network.

We take this a stage further with the orchestration of resources across distributed edge nodes with the aim of load balancing, improving resource utilisation, reduced response times \emph{and} increased throughput for user queries. With the above background and the security and privacy advantages of context-aware and timing-based data storage, we approach the subject of context-aware, similarity-based, orchestration of data and service management, towards better QoE and resource utilisation.

\subsection{Computation Deduplication and Reuse}


In cloud computing, the storage of several data duplicates becomes an important issue in terms of information distribution, storage reliability and security~\cite{ye2010secure}. 
Data deduplication has been previously researched in parallel with CDN systems~\cite{harnik2010side}, \cite{7040510} to increase the efficiency of data distribution and availability within clouds \cite{7161551}. Thus, research on data deduplication 
at the edge can benefit from prior research in cloud computing data deduplication, with the added advantage of utilising a readily distributed system.


The direction of data deduplication at the Edge of the Internet has lately gained traction 
~\cite{guo2018potluck}~\cite{9762397}. As a result, data and function reuse have also become more popular research subjects~\cite{10.1145/3609504}. Most techniques for data reuse assume that specific content is addressed by name~\cite{10.1145/2656877.2656887}, \cite{mastorakis2020icedge} or by a hash of the name \cite{guo2018potluck}. While \cite{guo2018potluck} considers data storage and reuse across applications it does not investigate the distribution of cached information across a network. Conversely, we access cached and stored results based on the similarity of the query rather than using a name as an index to a specific piece of content and we use the content's locality of reference as a better and faster identification, lookup and distribution mechanism. Further, metrics such as measurements of the above-mentioned are used towards better resource-and-service orchestration in our implementation.



Similarity has been previously approached from different points of view (mostly IoT and VR), to better implement and make use of computing at the edge~\cite{10.1145/3373376.3378516} and \cite{9488757}. On the other hand, these approaches only consider edge-based, end-user, heterogeneous devices, where the environment could be hard to control and secure. These processes can add complexity to the system, create fragmentation in resource allocation and imply a large overhead in processing. This can be an especially hard to solve problem when considering end-user devices' computing capabilities and availability.

\vspace{-0.2cm}

\subsection{Locality Sensitive Hashing and Network Services}

Locality Sensitive Hashing (LSH)~\cite{FALCONN,10.1145/2020408.2020578,10.5555/1325851.1325958,10.1145/1873951.1874168} is a technique that exploits hash collisions in such a way that similar queries are mapped to the same bucket. In \cite{10.1145/3447786.3456234} and \cite{9246572}, the authors use the capabilities of LSH to improve the way in which networks are managed, by \emph{improving routing tables and SDN placement through hashing entries} and using LSH for similarity-based routing, further \emph{improving network security} in the process, by implementing hash-based identity anonymisation.


Some works have previously used LSH with the network itself~\cite{9488833} or processing data/code/results (e.g.~\cite{guo2018foggycache} and \cite{9488722}), in order to \emph{improve processing resource usage and timings}. These applications for LSH are very common as of recent, in the research of computation systems. However, the context of these improvements has varied and was never considered to be a network- and/or storage-based system.

One of the most relevant and useful domains to use LSH is within storage environments, due to the ability of data (or certain features) to be hashed~\cite{rfc6920, saino2014icarus}, \cite{10.1145/2491224.2491232}. Further, data has the ability to be split into smaller ``chunks'', aggregated and/or deduplicated, in similar and/or purpose-specific conglomerates~\cite{10.1145/3373376.3378516}, forming more useful "pools", from which applications and tasks can draw any needed data as "prime materials". Studies into \emph{localised storage and computing systems} include \cite{10.1145/3410220.3460103}.

The advantage of local storage optimisation via LSH (LSH being applied \emph{externally or within the storage/computing system}), was previously used for in-network, general storage~\cite{9488757}, and computing \cite{guo2018foggycache}, towards achieving better system performance. What is not approached in the above works is session-less, network-based storage, and its implications. We take advantage of LSH in our work to realise the semantics of computation-context distribution and reuse.


\subsection{Use Cases}

As the IoT and Internet Edge environments evolve, the Edge includes more and more audio/video (A/V) devices, which can be used for increasingly diverse tasks and/or services needed throughout the Internet. There are so many IoT and "smart"/connected devices (e.g. CCTV cameras, drones, smart assistant speakers), that a single EDR (server) will not be enough. However, having multiple servers 
means that load balancing queries between servers needs to distribute requests in such a way to maximise the opportunity for reuse. Thus, we present the use cases of traffic cameras (among the video/image use cases) and personal assistants (such as Amazon Alexa, Google Assistant, for NLP and voice command interpretation) as our two example applications, going through the algorithm implemented within our \sol design.

As it happens, most of the Edge networks (with multi-access Edge computing) nowwadays have specific capabilities and can deal with certain types of use/test cases. Thus, certain edge domains (which can be more or less geographically distributed) can be specialised in many or only specific use cases (including the ones presented in this paper, as examples). In this work, we will assume that one domain provider can handle either of the two use cases.

We approach two use cases within both parts of our evaluation:


\begin{enumerate}[leftmargin=0cm,itemindent=.4cm,labelwidth=\itemindent,labelsep=0cm,align=left, noitemsep, topsep=0pt]
    \item Video frames/Images object detection/classification - Cameras are supplying captured video frames as queries to edge servers which are processing those frames to detect and count cars and return the number and types of cars in the image to the user. Such cameras may have frame rates ranging from a frame every $\sim$300ms, down to one per second. As mentioned above, one example function based on the video frames, deployed within the EDR (and its environment), may be to estimate the traffic volume for all the lanes, travelling in the same direction of an intersection. Consecutive camera frames, are offloaded to an EDR, so that the approximate number of cars is detected and the traffic volume is estimated. 
    
    In this case the video-frame similarity of the input queries can be very high, generally, considering the format of the files - the user supplied video frames, associated with processing by certain (e.g. counting, classification, detection) services. Data of this format can easily be reused at the edge, for low-latency response applications. 
    
    \bigskip
    
    \item Command identification from spoken input - Users are supplying recordings of voice commands, by accepting that the personal assistant shares encoded/hashed recordings, as queries to edge servers which process them to perform speech-to-text translation and text-to-service classification, for actuation, before returning the resulting service call to the user. 
    Smart assistant devices such as the above may have voice command generation rates ranging from an utterance every $\sim$5ms ($\sim$200 devices/EDR), down to one every 1-5s ($\sim$5-10 devices/EDR). In this example, an EDR-based function, taking the voice snippets as input, can infer an exact action, based solely on the snippet, rather than two other functions. 
    Voice requests can be offloaded to an EDR at any point in time, so that the command is assessed and the right triggers are put in place, in the local smart environment. 
    
    \indent In this case, there can be high similarity within the same voice profiles, comprised of encoded/hashed speech snippet datasets, provided by the smart-assistant(s), associated with the NLP service, based on previous speech-to-text classification and feature extraction. This kind of data is very well suited, to be encoded for reuse at the edge, in order to take advantage of locality and the privacy-preserving environment. 
    
\end{enumerate}
\bigskip


We used four datasets in total, two to represent different potential extremities of data present in each of our two use cases. These are: the MNIST dataset~\cite{MNIST_as_JPEG}, a traffic CCTV dataset, generated by one of the authors, placed next to a highway junction and capturing video, which is mainly meant for traffic detection around the junction, a dataset which contains anonymised, "Alexa" wake-word, voice snippets \cite{Alexa} and another, general smart-home voice commands dataset, comprised of anonymised voice commands \cite{General_Commands}

\section{System Design}
\label{sec:Overview}

\begin{figure}[tbh]
\vspace{-0.2cm}
\centering
\includegraphics[width=1\columnwidth]{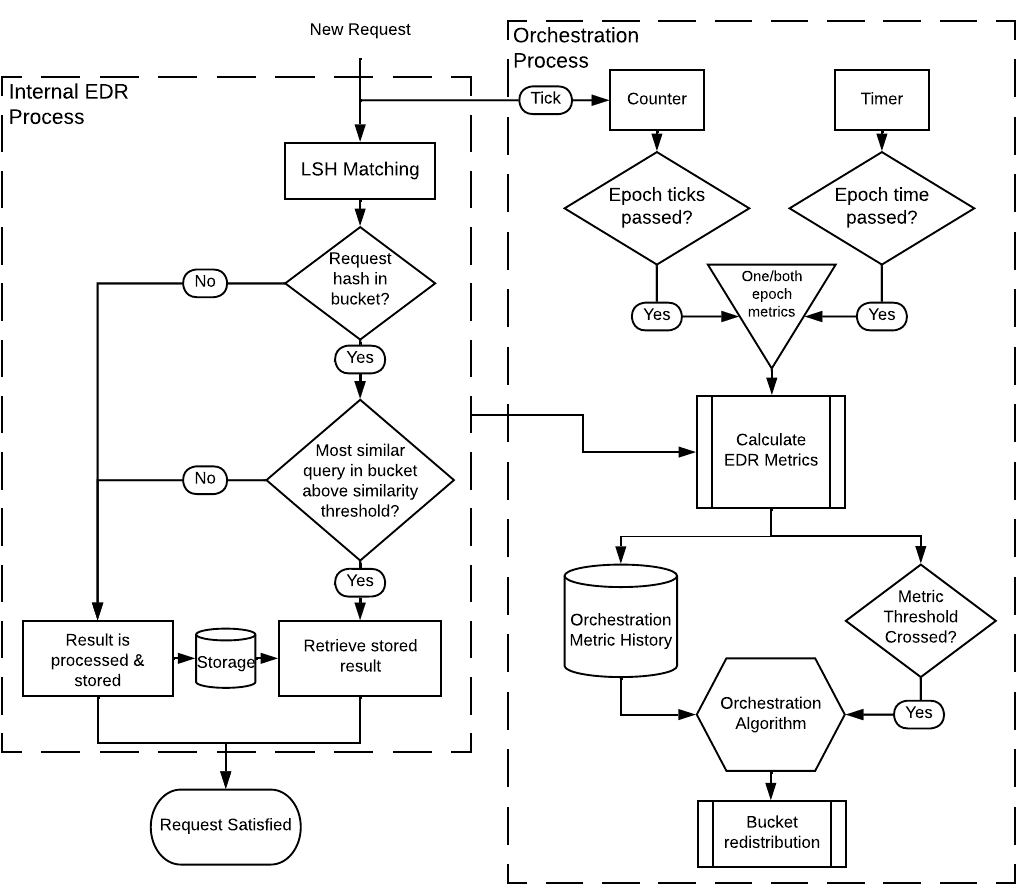}
\vspace{-0.2cm}
\caption{Overview of \sol Workflow}
\label{fig:ReStorEdge_Schematic}
\vspace{-0.2cm}
\end{figure}

Our design is based around using the above-mentioned, well-studied LSH mechanism \cite{Haghani2008LSHAL} for improving the performance of computation (reuse) in storage-enabled networked environments, and to satisfy the necessity of the similarity-based lookup that the envisioned system is using. Further, orchestration is implemented in order to maximise both system performance and service delivery efficiency. In Figure~\ref{fig:ReStorEdge_Schematic}, we present an overview of the \sol operational workflow. \sol includes a data storage and processing location management component 
- the orchestrator - which uses EDR-specific and system-wide metrics such as CPU-usage and data reusability to optimise the distribution of processing and storage amongst EDRs.
The other key component is the EDR environment which routes, processes, stores and redistributes queries and data through the system, as directed by the orchestrator. 

Looking at the system from a top-down perspective, we can provide a few black-box views, in order to simplify the explanation and provide readers with a modular understanding of the system, as follows:

\begin{itemize}[leftmargin=0cm,itemindent=.3cm,labelwidth=\itemindent,labelsep=0cm,align=left, noitemsep, topsep=0pt]
    \item From a high-level perspective, application-specific data and queries are generated at the edge, and they are processed by the system to return the results of the queries to the users in due time.
    \item Looking in more detail, but above the level of orchestration, it can be understood that the system achieves the above-defined performance by processing data, analysing query similarity, storing query identifiers and all the relevant data and reusing the relevant results from such queries, in an efficient way.
    \item The above-mentioned efficient way of processing/reusing requests and results, deploying the right functions (and data) in the right places, and achieving good QoS is done through EDR, data and function orchestration and the application of LSH techniques. These are the main subjects of this paper, and they will be explained in more detail through the remainder of this section.
\end{itemize}


LSH is used as it would be nearly impossible to find similarity between data pieces in a short enough time period in such a realistic environment (with large amounts of data) with the time complexity of normal, linear lookup solutions. Note that this short lookup time has to make it worth the lookup, when compared to reprocessing the respective queries. 

\vspace{-0.1cm}
\subsection{Definitions of terms}
\vspace{-0.1cm}

\textbf{LSH query/hashing process} - this process maps an incoming query to a bucket, which contains all prior queries with the same hash value. Thanks to the LSH algorithm all queries in the same bucket will be similar.



\textbf{Reuse} - a request/query causes a reuse HIT when the result of LSH (array of nearest-neighbour hashes), combined with the (two-process-) application-specific-similarity-check, return directly the results of a “similar-enough” query. A query causes a MISS when the above-mentioned process fails any of the checks. The MISS causes the query-associated function to go through the classic “query” or “processing” function.

\textbf{“Similar enough”} - this feature is application-dependent (or it can be EDR-provider-dependent, depending on priorities), and it is determined depending on metrics such as “needed” accuracy, system performance (EDR-environment-related) and/or latency constraints;

\textbf{Application-specific similarity check} - this “filters through” a nearest neighbour hash array (resulting from LSH), of (previously queried) same-application-bound query hashes.

\textbf{Satisfied request} - any kind of request that is received by the system, that is delivered on time and with a accuracy over the threshold imposed by its application-specific requirements (most probably settled via service-level agreements - SLAs) is defined as satisfied (whether through computation or reuse). \textbf{Note} that the satisfied requests might be less accurate/correct. However, if the requests are generally proven to be (historically) proven to be within a certain similarity threshold and their results relevant, they will be accurate \textit{enough} to satisfy any requests originating from the same (type/context of) application(s).\cite{9762397} 


\vspace{-0.1cm}
\subsection{Description of ReStorEdge Operational Workflow}
\vspace{-0.1cm}

As it can be seen in Fig. \ref{fig:ReStorEdge_Schematic}, after a new request enters the system, at the edge nodes (the first-hop EDR), it is processed in two different ways, by two different processes, while also being routed (and possibly stored, for different purposes, outside of the context of this study).

\begin{itemize}[leftmargin=0cm,itemindent=.3cm,labelwidth=\itemindent,labelsep=0cm,align=left, noitemsep, topsep=0pt]
    \item The main processing path is the internal EDR environment process. This includes most of the data routing, so for simplicity, we will not present routing as a separate process. We shall detail this process below:
    \begin{enumerate}[leftmargin=.3cm,itemindent=.4cm,labelwidth=\itemindent,labelsep=0cm,align=left, noitemsep, topsep=0pt]
        \item When the request enters the system, the application tag/label of the data is checked and LSH is applied for the hash to be generated;
        \item If the hash value of the current request identities a bucket of prior requests with the same hash value, the query becomes a reuse candidate (application-permitting, of course - which is why application association is the first check);
        \item If there are prior queries in the bucket and the current request data is similar enough with the most similar (LSH-determined) query data in the bucket, (stored) the results of the previous query data are looked up, fetched and returned to the appropriate user (if requested/subscribed).
        \item If the request data is dissimilar/not similar enough to previous queries, it is processed and the results are sent to both the appropriate output gateway, as well as to the local (processing EDR's) storage. The query data is stored along its results, for the same hash entry.
    \end{enumerate}

    \item The right-hand part of the workflow figure shows how the orchestration process is triggered periodically to optimise the assignment of buckets to EDRs.
    The process in this path relies more on the 
    the previous queries/previous system state, as detailed below:
    \begin{enumerate}[leftmargin=.3cm,itemindent=.4cm,labelwidth=\itemindent,labelsep=0cm,align=left, noitemsep, topsep=0pt]
        \item First, as there is an environment counter, counting ingress requests, it is incremented by 1, and the epoch ticks are counted. There also is an epoch timer. These are used for different types of orchestration strategies (or in conjunction);
        \item If the epoch has passed on this check (normally within at most a few ms), the following is done in parallel:
        \begin{itemize}
            \item A check is made, whether the metric threshold of the respective strategy is crossed;
            \item The orchestration metric history is updated;
        \end{itemize}
        \item Orchestration algorithm is called;
        \item Buckets are redistributed across the EDR Environment.
    \end{enumerate}
\end{itemize}

\vspace{-0.1cm}
\subsection{Query Routing}
\vspace{-0.1cm}

As noted in the operational workflow description, a new query is routed through the system as follows in these steps: (i) a device sends query data into the system, via its closest (local) EDR, as the first hop; (ii) the local EDR does LSH on the query, then looks up the hash in its forwarding table to find the EDR responsible for the associated bucket; (iii) the EDR responsible for the bucket takes the query and returns the result to the appropriate user (if requested/subscribed).

\emph{NOTE:} The forwarding tables in the appropriate EDRs, within the local environment, need to be updated whenever the orchestrator is triggered and reassigns buckets between EDRs.

\vspace{-0.1cm}
\subsection{Storage Similarity and Reuse}
\vspace{-0.1cm}

Once more data enters an EDR, the history of (application-specific) related data and information can be used in the advantage of the application \emph{and} system. Similarity, based on LSH and the new data's related information and meta-information, helps determine the (re)usability of existing query, function and associated resulting data.\footnote{EDR-based \sol prototype: \href{https://github.com/Chrisys93/LSH-SEND.git}{LSH-SEND.git}} 

\vspace{-0.1cm}
\subsection{Orchestration Algorithms}
\label{Alg}

\vspace{-0.1cm}
The problems of resource efficiency/utilisation, orchestration and load balancing are different when taken separately, but putting them together and considering the decentralised system behind them creates trade-offs and even some complexity problems. Thus, the placement of certain buckets in specific EDRs/areas of the EDR environment can be advantageous for many reasons, among which load balancing and resource utilisation optimisation in some cases, due to the predominance of data/functions of a more relevant type, topic, application, provider and/or data similarity in these EDRs or this area. Thus, for the system's service efficiency to increase, an orchestration algorithm is needed in certain cases, in order to optimise bucket placement. Such an algorithm could potentially be easily adapted and placed in a MANO-enabled, SDN-based system, as one of the main implementations of a system like ours.

We now introduce four orchestration algorithms (in Algorithms \ref{alg:queue}, \ref{alg:proc},  \ref{alg:proc_workload} and \ref{alg:proc_reuse}) 
that trade-off system resource usage and QoS. However, another trade-off comes up in the best case scenarios, and that is the trade-off between \emph{performance increase in both QoS and resource usage} and \emph{the cost for more strictly and intelligently managing the system over time}. The demonstration of these strategies will be evaluated with different datasets (representing different use cases) in section (\ref{sec:eval}), highlighting the above-mentioned, resulting trade-off.

Before execution, after a number of requests (epoch), a strategy trigger is checked and it does not reset the request count until the strategy triggers. This is done in all strategies, except CPU-Workload, where the strategy triggers on every epoch.


\begin{algorithm}\captionsetup{labelfont={sc,bf}, labelsep=newline}
\caption{Triggered, Queue-delay-based Orchestration Strategy}
\begin{algorithmic} [1]
\raggedright
\State \textbf{Inputs:} Input $i$ received from a user device
\If {Max\_queue\_delay($current\_node$) $>$ 2 $*$ $mean\_req\_proc\_time$}
\State $h\_d$ $=$ highest\_queue\_delay($no\_of\_buckets$)
\State $l\_d$ $=$ lowest\_queue\_delay($no\_of\_buckets$)
\EndIf

\For{($n\_h$, $h$, $n\_l$, $l$ in zip($h\_d$[0], $h\_d$[1], $l\_d$[0], $l\_d$[1]))}:
    \State self.controller.move\_bucket($n\_h$, $n\_l$, $h$, $l$)
\EndFor

\State distribute($i$, bucket\_update)
\State distribute($i$, bucket\_data\_update)
\State function\_statistics $fs$ $=$ update\_function\_statistics($i$)
\end{algorithmic}
\label{alg:queue}
\end{algorithm}

\begin{algorithm}\captionsetup{labelfont={sc,bf}, labelsep=newline}
\caption{Triggered, CPU-usage-based Orchestration Strategy}
\begin{algorithmic} [1]
\raggedright
\State \textbf{Inputs:} Input $i$ received from a user device

\If {CPU\_usage($current\_node$) $>$ $trigger\_threshold$}
\For{($N_{EDRs}$)}:
    \State $n\_h$, $h$ $=$ most\_proc\_usage()
    \State $n\_l$, $l$ $=$ least\_proc\_usage()
    \State move\_bucket($n\_h$, $n\_l$, $h$, $l$)
    \State move\_bucket\_data($n\_h$, $n\_l$, $h$, $l$)
    \State update\_orchestration\_CPU($n\_h$, $n\_l$, $h$, $l$)
\EndFor
\EndIf

\State function\_statistics $fs$ $=$ update\_function\_statistics($i$)
\end{algorithmic}
\label{alg:proc}
\end{algorithm}

\begin{algorithm}\captionsetup{labelfont={sc,bf}, labelsep=newline}
\caption{Epoch-based, CPU-Workload Orchestration Strategy}
\begin{algorithmic} [1]
\raggedright
\State \textbf{Inputs:} Input $i$ received from a user device

\If {$Request\_ticks$ $>$ $Epoch\_ticks$}
\For{($N_{Buckets}$)}:
    \State $n\_h$, $h$ $=$ most\_workload\_perBucket\_at\_update()
    \State $n\_l$ $=$ least\_orchestration\_workload\_perEDR()
    \State move\_bucket($n\_h$, $n\_l$, $h$)
    \State move\_bucket\_data($n\_h$, $n\_l$, $h$)
    \State update\_orchestration\_Workload($n\_h$, $n\_l$, $h$)
\EndFor
\EndIf

\State function\_statistics $fs$ $=$ update\_function\_statistics($i$)
\end{algorithmic}
\label{alg:proc_workload}
\end{algorithm}

\begin{algorithm}\captionsetup{labelfont={sc,bf}, labelsep=newline}
\caption{Epoch-based, CPU-triggered, CPU and Reuse Bucket Change Orchestration Strategy}
\begin{algorithmic} [1]
\raggedright
\State \textbf{Inputs:} Request $i$ received from a user device


\If {CPU\_usage($current\_node$) $>$ $trigger\_threshold$}

\For{($N_{EDRs}$)}:
    \State $n\_h$, $hp$ $=$ most\_proc\_usage()
    \State $n\_hr$, $hr$ $=$ most\_reuse()
    \State move\_bucket($n\_hp$, $n\_hr$, $hp$, $hr$)
    \State move\_bucket\_data($n\_h$, $n\_l$, $h$, $l$)
    \State update\_orchestration\_CPU($n\_hp$, $n\_hr$, $hp$, $hr$)
\EndFor

\For{($N_{EDRs}$)}:
    \State $n\_lr$, $lr$ $=$ lowest\_reuse()
    \State $n\_lp$, $lp$ $=$ least\_proc\_usage()
    \State move\_bucket($n\_lr$, $n\_lp$, $lr$, $lp$)
    \State move\_bucket\_data($n\_h$, $n\_l$, $h$, $l$)
    \State update\_orchestration\_CPU($n\_lr$, $n\_lp$, $lr$, $lp$)
\EndFor

\EndIf

\State distribute($i$, bucket\_update)
\State distribute($i$, bucket\_data\_update)
\State function\_statistics $fs$ $=$ update\_function\_statistics($i$)
\end{algorithmic}
\label{alg:proc_reuse}
\end{algorithm}




\section{Evaluation}
\label{sec:eval}


\subsection{Setup and Assumptions}

The evaluation is done in two phases. We first approach a local, prototype implementation, of the LSH-similarity-storage/processing system, and then we look into scaling the system up, to account for the whole EDR environment. The edge environment we consider in our evaluation is either a smart home or a (logically separated/network sliced) traffic-based environment/smart city, 
and the assumptions of the underlying systems and functions/storage services in place.

\textbf{\emph{Assumptions made on reuse}} - In this work, while we base our findings on realistic datasets, for the sake of computational resources and time efficiency, but with little loss of accuracy (an overall higher reusability), we base our simulation results on one reuse rating per bucket, instead of dynamically-varying reusability ratings in the simulations (this would require increased software and/or hardware complexity and would increase setup times).

\textbf{\emph{Assumptions made on processing resources}} - An important assumption we make, for simplifying system complexity is that the CPU cores do not use pipelining principles. However, these simplifications should not affect the overall performance evaluation in any positive manner.


\textbf{\emph{Assumptions made on realistic workload distribution and considerations for fair comparison}} - In the EDR environment simulator, while we take the realistic prototype evaluation outputs as inputs, we do a random distribution of associated requests and load across the system's ingress EDRs.


\subsection{EDR-based \sol Prototype Setup and Metrics}

For this setup we used FALCONN \cite{FALCONN} for all LSH queries on the datasets, the "tiny" version of YOLO object detection \cite{YOLO} for the image processing software and pyAudioAnalysis \cite{pyAudio} and pocketsphinx \cite{1659988}, as audio feature detection and classification tools. We implemented this on an Intel i7-8650U, 16GB DDR4, 512GB NVMe SSD system, via Ubuntu 20.04.

This is a prototype of what needs to be implemented at scale, for the system to provide storage, processing and reusability. This implementation provides the means for the algorithms introduced in section \ref{sec:Overview} to be evaluated. Further, this evaluation step provides realistic metrics and adds to the accuracy, credibility and realistic outlook of the second (simulation) part of the evaluation (\ref{Sim_Eval}). Thus, the main two metrics (and determining factors, for the simulation-based evaluation) are shown below.

\textbf{\emph{Performance Indicators}}

\begin{itemize}[leftmargin=0cm,itemindent=.3cm,labelwidth=\itemindent,labelsep=0cm,align=left, noitemsep, topsep=0pt]
    \item \emph{General Reusability of the datasets} - This metric provides us with a preview of the amount we would expect data of certain types, topics and from certain sources to be reused, on average. We use this as a metric to determine the way in which the influence of certain orchestration approaches and resource allocation would impact system performance.
    \item \emph{Overall request satisfaction times} - The trade-off implied  between reuse and processing can be less obvious in a networked system, where we have to also take internal system delays and network delays into account. Thus, we take previous measurements, and see the difference between a request-lookup-execute-return and a request-match-lookup-fetch-return processes.
\end{itemize}


\subsection{EDR-based \sol Prototype Evaluation}
\label{LSH_Eval}

\begin{figure}[h]
    \centering
    \subfigure[MNIST]{
        \includegraphics[width=0.45\columnwidth]{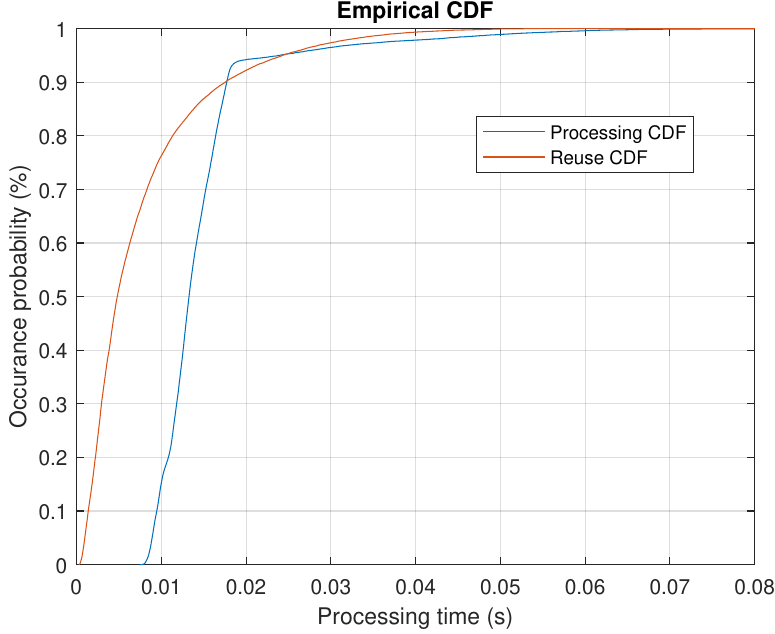}
        \centering
        \label{subfig:CDF-proc_MNIST}
    }
    \subfigure[Traffic Detection]{
        \includegraphics[width=0.45\columnwidth]{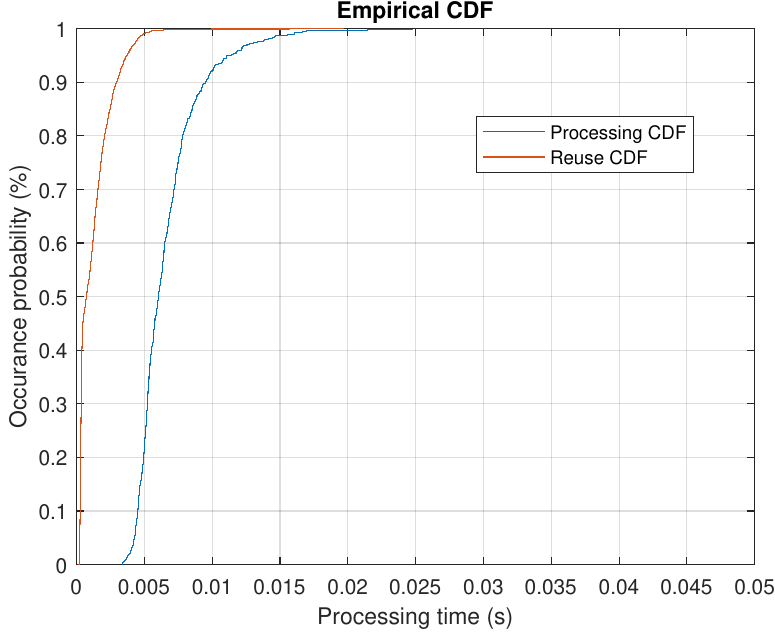}
        \centering
        \label{subfig:CDF-proc_Traffic}
    }
    
    \subfigure[Alexa]{
        \includegraphics[width=0.45\columnwidth]{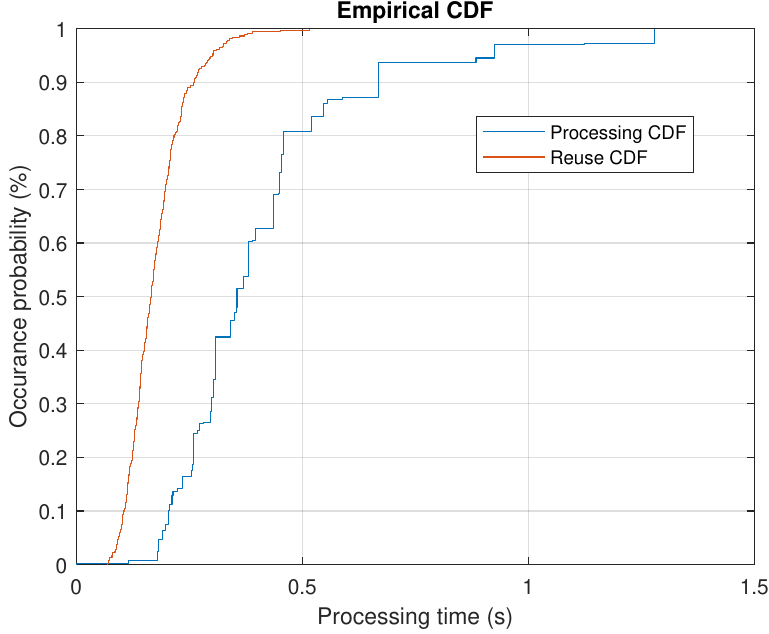}
        \centering
        \label{subfig:CDF-proc_Alexa}
    }
    \subfigure[General Commands]{
        \includegraphics[width=0.45\columnwidth]{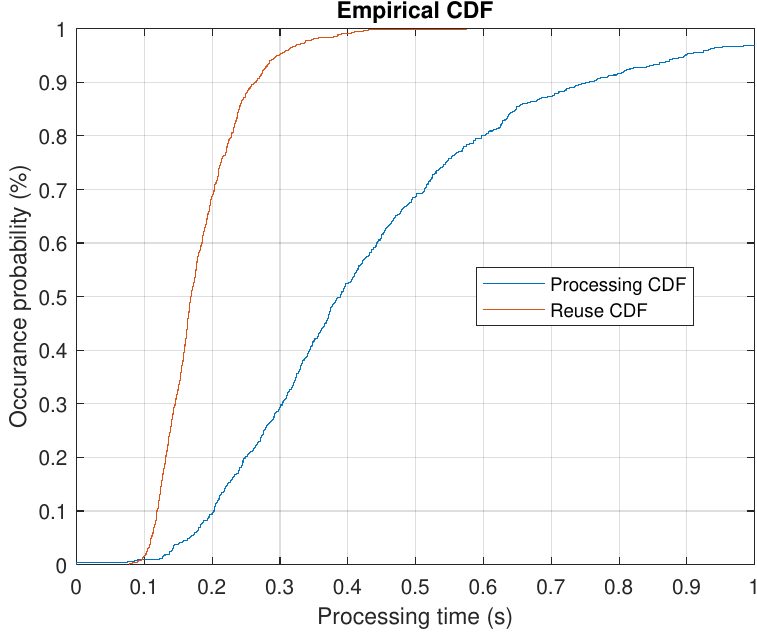}
        \centering
        \label{subfig:CDF-proc_General}
    }
    \caption{Processing vs. Reuse timings for different datasets.}
    \label{fig:proc-CDFs}
\end{figure}

\emph{Processing time improvements} - One of the biggest trade-offs we are clarifying in the case of these results are of effective information processing timings. It is \emph{essential} that information is distributed, stored, processed and consumed as efficiently and timely as possible. Thus, looking at the plots of processing/reuse CDFs of the image use cases (subfigs \ref{subfig:CDF-proc_MNIST}, \ref{subfig:CDF-proc_Traffic}), we can see a definite trade-off between function processing- and reuse-associated delays. A first (and the most obvious) observation is that, the larger the processing times, the more important reuse is for minimising latency. However, a more important trade-off is that, with the appropriate orchestration, the top of both curves can be used (subfig \ref{subfig:CDF-proc_MNIST}), to the advantage of both application providers and the EDR service provider. This will become more obvious in the next subsections' analysis of the CPU-usage and CPU-Workload srategies' impact on the system, in simulations.

\emph{Processing times and LSH searches} - While LSH and reuse help by fetching the needed data faster than processing the similar data, there are still other processes that slow the fetch down. However, for a good service and a more efficient system, these processes have to be maintained. For the case of sound files, the above-mentioned process is represented by feature extraction, which, while efficient, it hinders the fast Nearest Neighbours search that LSH entails. Making a parallel comparison, between the image and audio analysis cases, we can see a large increase in LSH timings (NN search timing) from the audio to the image datasets (LSH timing Plots/Table \ref{tab:reuse_results}), however, we can see a large (but proportional) decrease in processing times in the same direction. Thus, comparing the processing plots in Figure \ref{fig:proc-CDFs} and the LSH timing in Table \ref{tab:LSH_timings}, we can conclude that, as pre-processing is done regardless of whether queries are just processed from scratch or they trigger reuse, the main importance in the matter of timings should be attributed to whether reuse is/should be done or not.

\setlength\tabcolsep{1pt} 
\begin{table}[h]
\caption{LSH Nearest Neighbours Search Timings Table}
\label{tab:LSH_timings}
\resizebox{\columnwidth}{!}{\begin{tabular}{|c|cccc|}
\hline
\multirow{2}{*}{\textbf{\begin{tabular}[c]{@{}c@{}}Similarity\\ Threshold\end{tabular}}} & \multicolumn{4}{c|}{\textbf{Datasets LSH nearest Neighbour Search Avg (ms)}} \\ \cline{2-5} 
     & \multicolumn{1}{c|}{MNIST}  & \multicolumn{1}{c|}{Traffic Detection} & \multicolumn{1}{c|}{Alexa}    & General Commands \\ \hline
60\% & \multicolumn{1}{c|}{5.3} & \multicolumn{1}{c|}{0.6}            & \multicolumn{1}{c|}{0.1} & 0.14          \\ \hline
70\% & \multicolumn{1}{c|}{6.8} & \multicolumn{1}{c|}{0.6}            & \multicolumn{1}{c|}{0.1} & 0.14          \\ \hline
80\% & \multicolumn{1}{c|}{6.8} & \multicolumn{1}{c|}{0.8}            & \multicolumn{1}{c|}{0.1} & 0.14          \\ \hline
90\% & \multicolumn{1}{c|}{6.9} & \multicolumn{1}{c|}{0.9}            & \multicolumn{1}{c|}{0.1}  & 0.15          \\ \hline
\end{tabular}}
\end{table}

\emph{Data Reusability} - Since we are interested in the reuse of processed data (for different reasons/applications), we want to see \emph{what applications} can rely more or less on reuse, \emph{when} they can rely on processing reuse and \emph{how much reuse needs to be taken into account}, to obtain the best results. Further, since certain data types and/or topics may be more or less correlated to each other, it may be that some similarity thresholds and application-specific associations need to be used, to "filter down" most of, if not all, false-positive reused results. Looking at the MNIST part of the Reuse Table (Table \ref{tab:reuse_results}), we can see a large difference in the reusability of data from the 60\% similarity to the 70\% thresholds. That is due to the big difference in the amount of change expected, having a high amount of space covered strictly by the number in the images of the MNIST dataset.

\setlength\tabcolsep{1pt} 
\begin{table}[h]
\caption{LSH Reusability Table}
\label{tab:reuse_results}
\resizebox{\columnwidth}{!}{\begin{tabular}{|c|cccc|}
\hline
\multirow{2}{*}{\textbf{\begin{tabular}[c]{@{}c@{}}Similarity\\ Threshold\end{tabular}}} &
  \multicolumn{4}{c|}{\textbf{Datasets Reusability Avg (\%)}} \\ \cline{2-5} 
 &
  \multicolumn{1}{c|}{MNIST} &
  \multicolumn{1}{c|}{Traffic Detection} &
  \multicolumn{1}{c|}{Alexa} &
  General Commands \\ \hline
60\% & \multicolumn{1}{c|}{12}   & \multicolumn{1}{c|}{70}   & \multicolumn{1}{c|}{86.6} & 18.6  \\ \hline
70\% & \multicolumn{1}{c|}{0.74} & \multicolumn{1}{c|}{72.5} & \multicolumn{1}{c|}{81.8} & 18.4  \\ \hline
80\% & \multicolumn{1}{c|}{0.42} & \multicolumn{1}{c|}{67.2} & \multicolumn{1}{c|}{59.3} & 18.3  \\ \hline
90\% & \multicolumn{1}{c|}{0.26} & \multicolumn{1}{c|}{60.8} & \multicolumn{1}{c|}{25.3} & 19.36 \\ \hline
\end{tabular}}
\end{table}

\subsection{\sol Environment Simulator Setup and Metrics}

Now we have realistic measurements of a localised example and we can expand the model of the system, in order to implement orchestration strategies, and to make the case for a more realistic, decentralised and coordinated system. With these, we can now scale our experiments up and look at an EDR-based simulation, for which the impact of the orchestration strategies will be critically analysed and assessed.

\textbf{\emph{Orchestration Metrics and Evaluation Considerations}}

Considering the environment of a smart city, users will interact with applications/services hosted on edge nodes by forwarding queries. We consider EDRs that are distributed over the edge/access network and are able to process requests and cache results of prior queries. Through this part of the evaluation we demonstrate the efficiency of the data placement algorithms, show superior QoS and resource allocation performance and consider all the imposed trade-offs. Finally, we show how the system could be fine-tuned, for the best outcomes (depending on the situation it is put in and the specific application/service providers' demands), through software simulations. 

\emph{NOTE:} There is a big difference between the hardware resources used for realistic simulations and a hardware-implemented prototype, which makes the hardware implementation more desirable, directly applicable and dedicated. However, a hardware implementation would make the evaluation process slower, more complex and costly, while making results generation more reliable, realistic and faster.


\setlength\tabcolsep{1pt} 
\begin{table}[h]
\caption{Simulation parameters and used datasets.}
\label{tab:params}
\resizebox{\columnwidth}{!}{%
\begin{tabular}{|c|c|}
\hline
\textbf{Parameter} &
  \textbf{Value(s)} \\ \hline
Number of EDRs &
  15 \\ \hline
\begin{tabular}[c]{@{}c@{}}Distance and link delay\\ between EDRs at the\\ edge level\end{tabular} &
  \begin{tabular}[c]{@{}c@{}}2ms following values reported by\\ recent studies \cite{bgp}\end{tabular} \\ \hline
\multirow{4}{*}{\begin{tabular}[c]{@{}c@{}}Used\\ Datasets\end{tabular}} &
  \begin{tabular}[c]{@{}c@{}}MNIST: 42,000 \\ unique characters \\ Workload: 250-500-1000 reqs/s \end{tabular} \\ \cline{2-2} 
 &
  \begin{tabular}[c]{@{}c@{}}Self-created CCTV Traffic Camera\\ Feed Dataset: 3000 Video Framess \\ Workload: 2000-4000-6000 reqs/s \end{tabular} \\ \cline{2-2} 
 &
  \begin{tabular}[c]{@{}c@{}}Alexa Wake-word \\ Dataset: 365 Audio Filess \\ Workload: 250-400-800 reqs/s \end{tabular} \\ \cline{2-2} 
 &
  \begin{tabular}[c]{@{}c@{}}Smart-Home General Command\\ Dataset: 899 Audio Filess \\ Workload: 150-200-300 reqs/s \end{tabular} \\ \hline
Data Sizes &
  \begin{tabular}[c]{@{}c@{}}Ranging from 40KB, that represent\\ MNIST letter images, to 5MB, that\\ represent high-quality video frames\end{tabular} \\ \hline
Link delay between EDRs &
  2ms \\ \hline
\begin{tabular}[c]{@{}c@{}}Link delay between a\\ gateway and its \\ adjacent EDR\end{tabular} &
  2ms \\ \hline
\end{tabular}%
}
\end{table}



    \setlength\tabcolsep{1pt} 

\textbf{\emph{Performance Indicators}}
\begin{itemize}[leftmargin=0cm,itemindent=.3cm,labelwidth=\itemindent,labelsep=0cm,align=left, noitemsep, topsep=0pt]
    \item \emph{Throughput} - number of processed messages per second - its main modifiers are as follows:
    \begin{itemize}
        \item Sum of (built up) RTT delays;
        \item Dataset reusability vs. system processing capacity
    \end{itemize}
    \item \emph{Resource utilisation} - amount of processing power and network resources, consumed to maintain good QoS and/or QoE - the main modifiers of this indicator are as follows:
    \begin{itemize}
        \item Edge- or Reuse-satisfied requests;
        \item Processing load distribution;
    \end{itemize}
    \item \emph{Number of orchestration calls} - done for each algorithm, during each run, and shows how efficient the management system is in doing its job.

\end{itemize}

\subsection{\sol Environment Simulations Evaluation}
\label{Sim_Eval}

\begin{figure*}[h]
    \centering
    \subfigure
    {
        \includegraphics[width=0.3\textwidth]{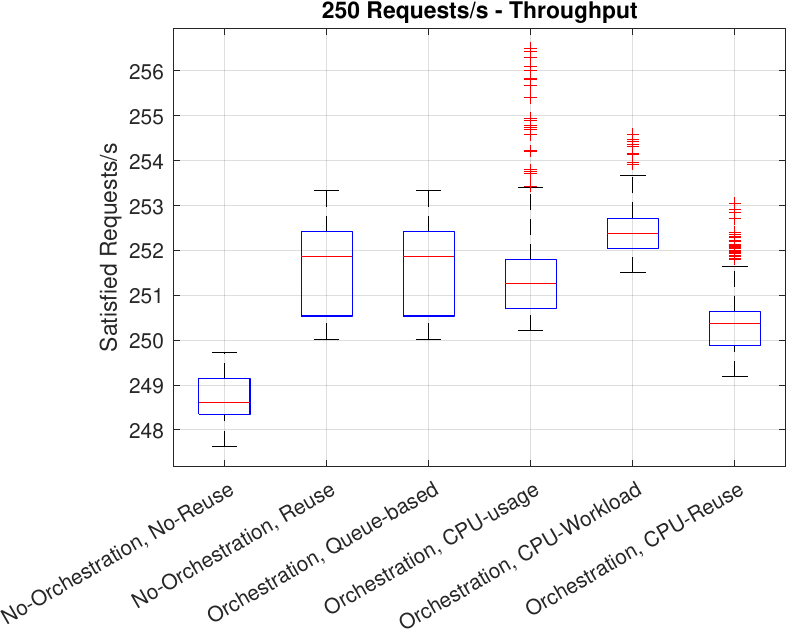}
        \centering
        \captionsetup{justification=centering}
        \centering
        \label{subfig:250_Throughput_effective_MNIST}
    }
    ~
    \subfigure
    {
        \includegraphics[width=0.3\textwidth]{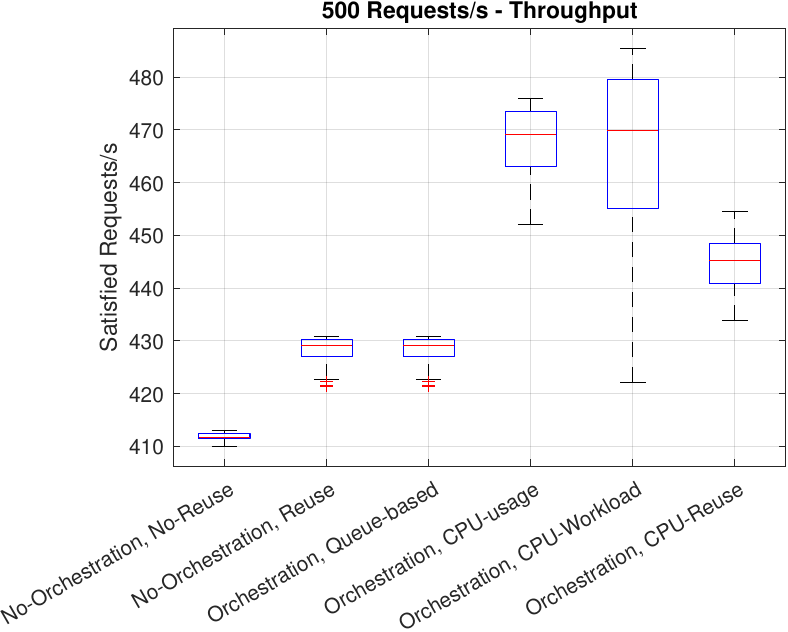}
        \centering
        \captionsetup{justification=centering}
        \label{subfig:throughput500_MNIST}
    }
    ~
    \subfigure
    {
        \includegraphics[width=0.3\textwidth]{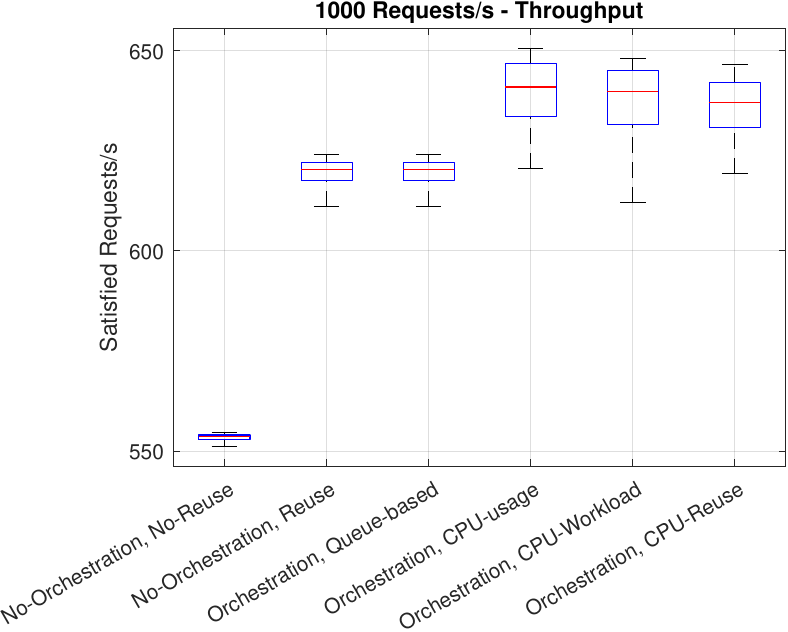}
        \centering
        \captionsetup{justification=centering}
        \label{subfig:throughput1000_MNIST}
    }
\vspace{-0.2cm}
\caption{Throughputs for different rates with MNIST Dataset.}
\label{fig:throughputs_MNIST}\vspace{-0.1cm}
\end{figure*}

\begin{figure*}[h]
\centering
\subfigure{
\includegraphics[width=0.3\textwidth]{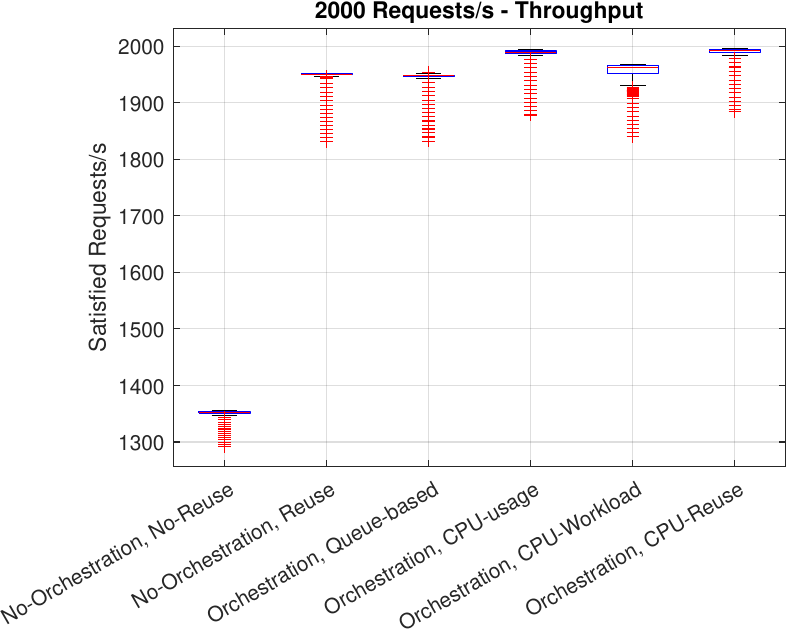}
\centering
\centering
\label{subfig:throughput2000_Traffic}
}
~
\subfigure{
\includegraphics[width=0.3\textwidth]{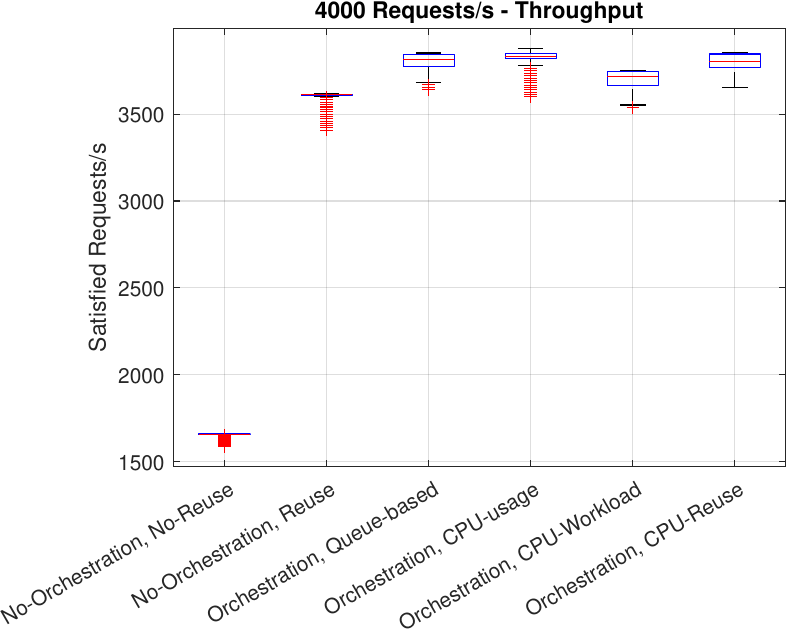}
\centering
\label{subfig:throughput4000_Traffic}
} 
~
\subfigure{
\includegraphics[width=0.3\textwidth]{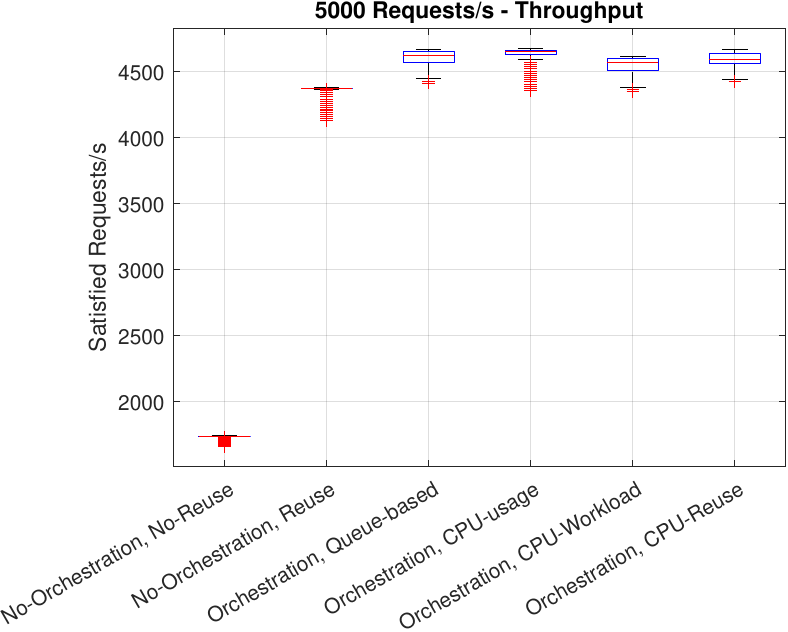}
\centering
\label{subfig:throughput5000_Traffic}
}
\vspace{-0.2cm}
\caption{Throughputs for different rates with Traffic Detection Dataset.}
\label{fig:throughputs_Traffic}\vspace{-0.1cm}
\end{figure*}

\begin{figure*}[h]
\centering
\subfigure{
\includegraphics[width=0.3\textwidth]{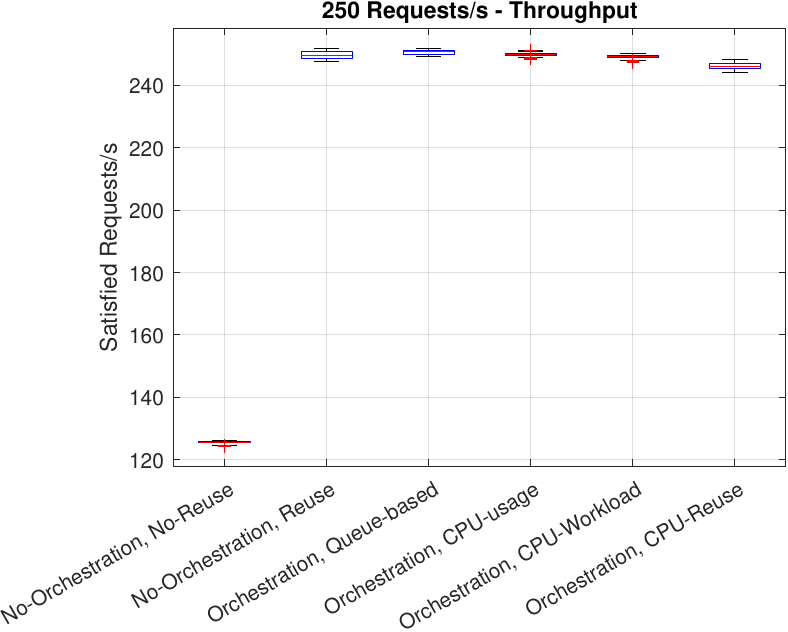}
\centering
\captionsetup{justification=centering}
\centering
\label{subfig:250_Throughput_effective_Alexa}
}
\subfigure{
\includegraphics[width=0.3\textwidth]{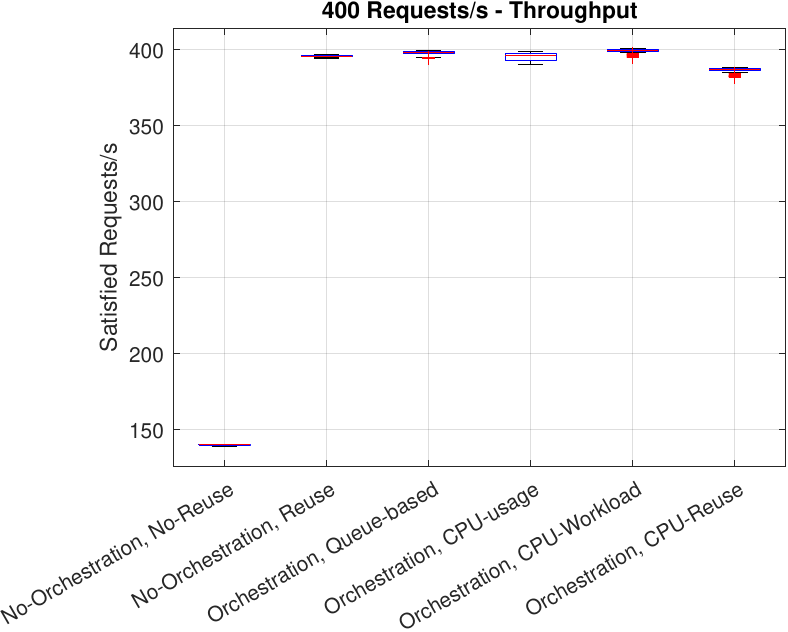}
\centering
\captionsetup{justification=centering}
\label{subfig:throughput400_Alexa}
} 
\subfigure{
\includegraphics[width=0.3\textwidth]{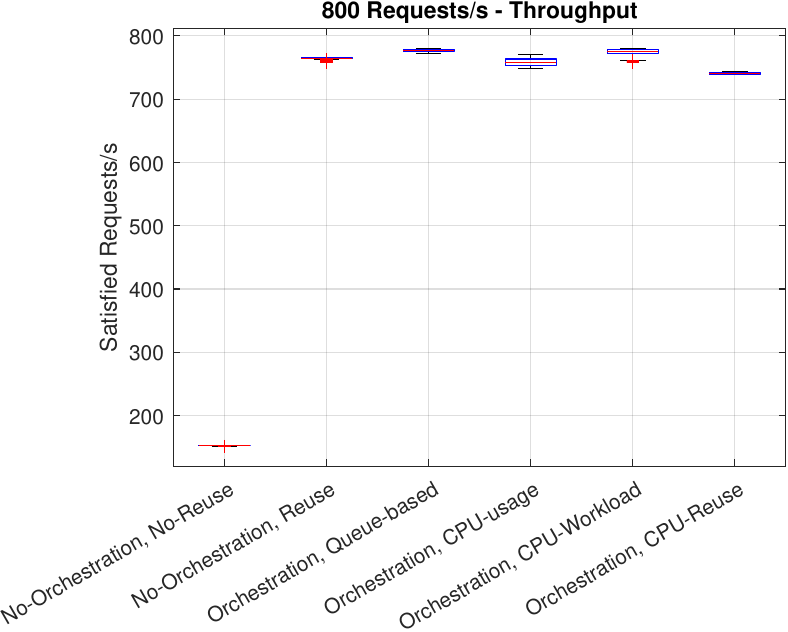}
\centering
\captionsetup{justification=centering}
\label{subfig:throughput800_Alexa}
}
\vspace{-0.2cm}
\caption{Throughputs for different rates with Alexa Dataset.}
\label{fig:throughputs_Alexa}
\vspace{-0.1cm}
\end{figure*}

\begin{figure*}[h]
    \centering
    \subfigure{
        \centering
        \includegraphics[width=0.3\textwidth]{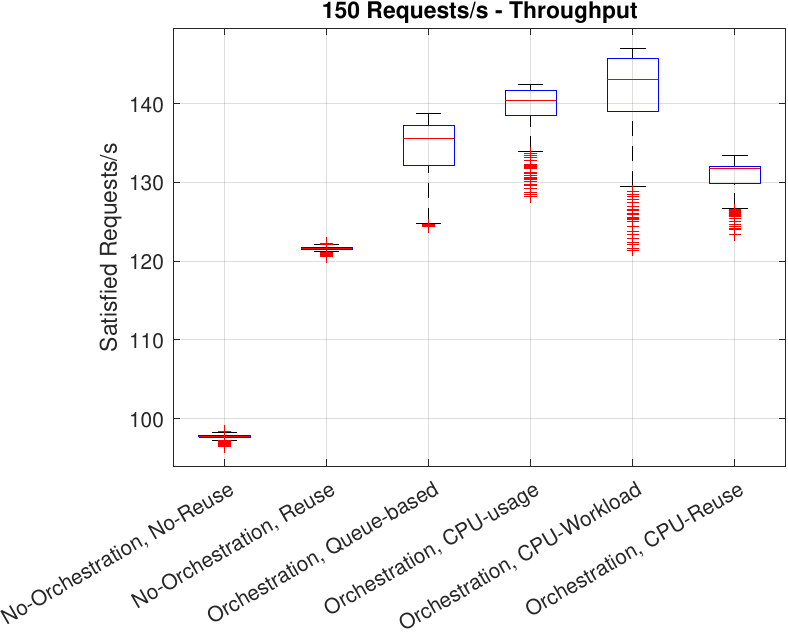}
        \centering
        \captionsetup{justification=centering}
        \centering
        \label{subfig:throughput150_General}
    }
    ~
    \subfigure{
        \includegraphics[width=0.3\textwidth]{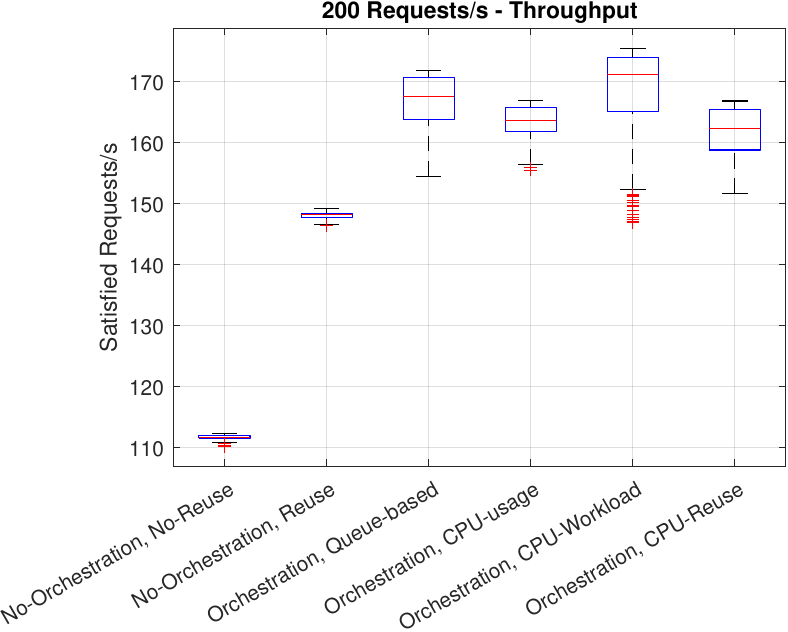}
        \centering
        \captionsetup{justification=centering}
        \label{subfig:throughput200_General}
    } 
    ~
    \subfigure{
        \includegraphics[width=0.3\textwidth]{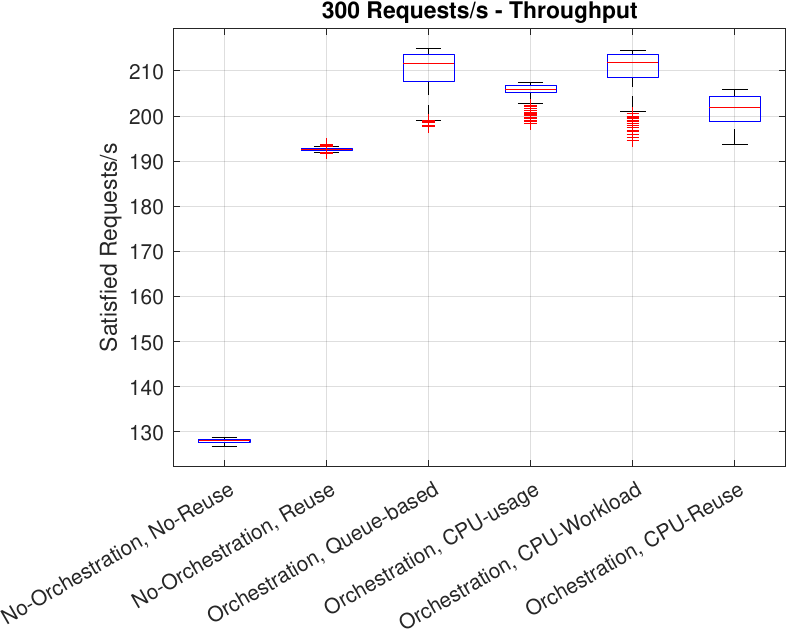}
        \centering
        \captionsetup{justification=centering}
        \label{subfig:throughput300_General}
    }
    \vspace{-0.2cm}
    \caption{Throughputs for different rates with General Commands Dataset.}
    \label{fig:throughputs_General}\vspace{-0.1cm}
\end{figure*}

The main purpose for this part of the evaluation is to show the diversity of data and the most appropriate approaches for different data types, topics and situations, depending on system-, user-, provider- and environment-defined variables. 

\emph{Throughput} - We shall start by comparing throughput and unprocessed messages at the end of simulations, among the different datasets and the applied orchestration strategies. Looking at the throughput of the different orchestration strategy case results obtained for the MNIST dataset (Figure \ref{fig:throughputs_MNIST}), we can see from the first look that the CPU utilisation strategy performs best among all orchestration strategies and across use cases.
By looking at the dataset, which has a large number of small sized queries and results, we can deduce the main reason for the aforementioned conclusion. In the traffic detection dataset, the throughput provided by both the queue-based and the CPU-reuse-based strategies have a (mostly) positive impact (Figure \ref{fig:throughputs_Traffic}), with the CPU-Reuse strategy providing continuous improvement to the no-orchestration, reuse case. The CPU-based negative impact is motivated by the strategy's concentration on CPU utilisation and lack of overall data reusability awareness, making the orchestration algorithm narrow-sighted. In high-reusability datasets, this is predominant. This shortcoming can be easily overcome at a later stage in system development, by more intelligently stopping any counter-productive orchestration. 
The throughput of the CPU-Reuse strategy in the case of the Traffic Detection dataset matches the performance of the queue-based and CPU-usage strategies in all workload cases (Figure \ref{fig:throughputs_Traffic}). While the throughputs of the Alexa dataset demonstrate a good overall performance of the orchestration strategies when compared to the no-orchestration, no-reuse case, due to the simplicity of the dataset, in the general, smart-home commands dataset (which is slightly more complex, both in utterance and command complexity and diversity), the CPU-Workload orchestration strategy provides the best performance improvements overall (Figure \ref{fig:throughputs_General}), detaching itself in the measurements of performance indicators, even though the Queue-based strategy may perform better in some cases/settings.


\begin{figure*}[h]
\centering
\subfigure{
\includegraphics[width=0.3\textwidth]{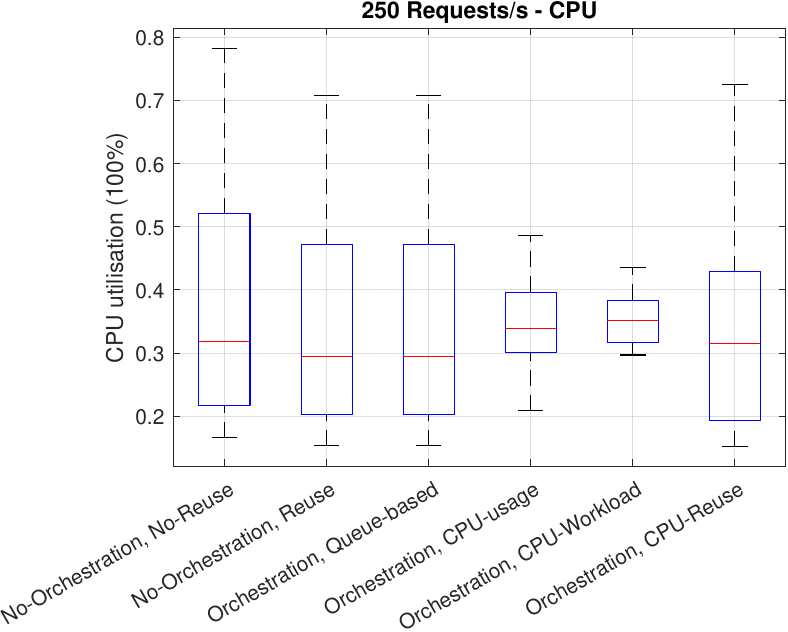}
\centering
\captionsetup{justification=centering}
\centering
\label{subfig:250_CPU_Avg_Box_effective_MNIST}
}
~
\subfigure{
\includegraphics[width=0.3\textwidth]{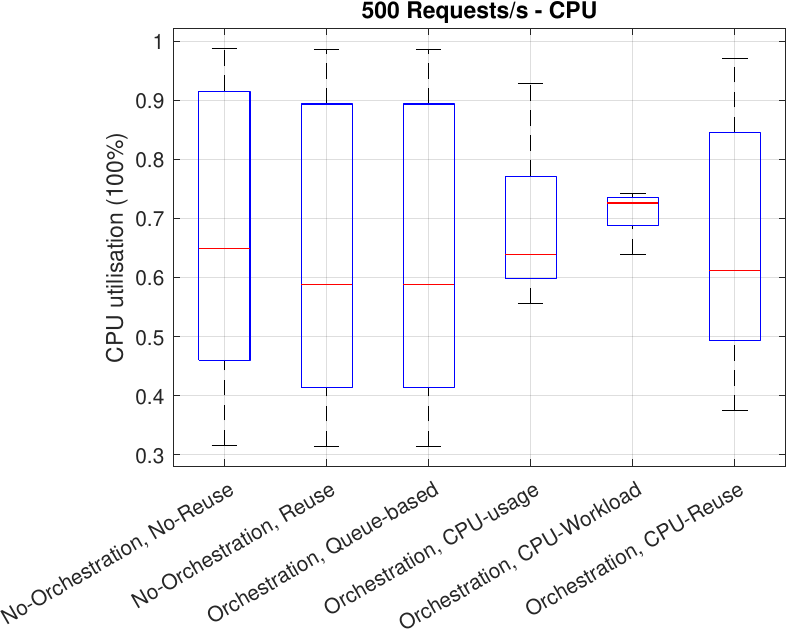}
\centering
\captionsetup{justification=centering}
\label{subfig:500_CPU_Avg_Box_effective_MNIST}
} 
~
\subfigure{
\includegraphics[width=0.3\textwidth]{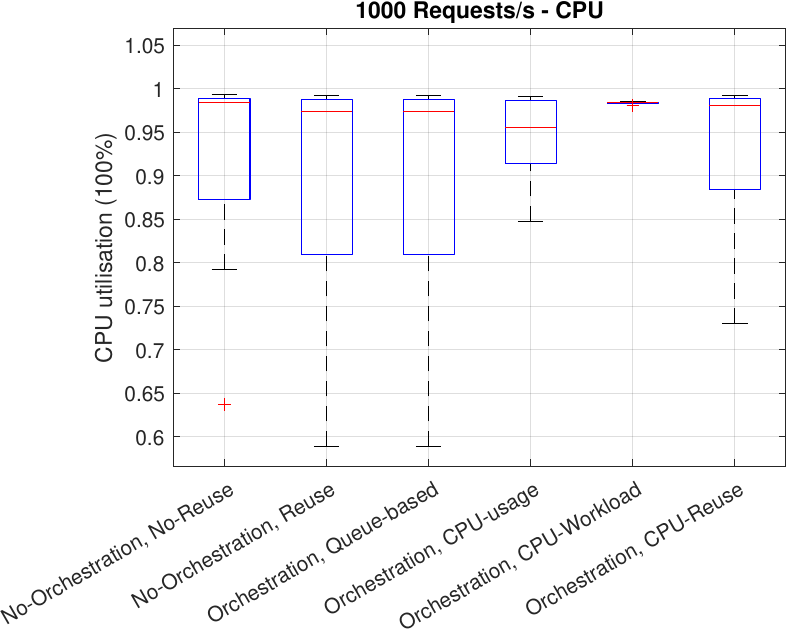}
\centering
\captionsetup{justification=centering}
\label{subfig:1000_CPU_Avg_Box_effective_MNIST}
}
\vspace{-0.2cm}
\caption{CPU utilisation for different rates with MNIST Dataset.}
\label{fig:CPU_MNIST}\vspace{-0.1cm}
\end{figure*}

\begin{figure*}[h]
\centering
\subfigure{
\includegraphics[width=0.3\textwidth]{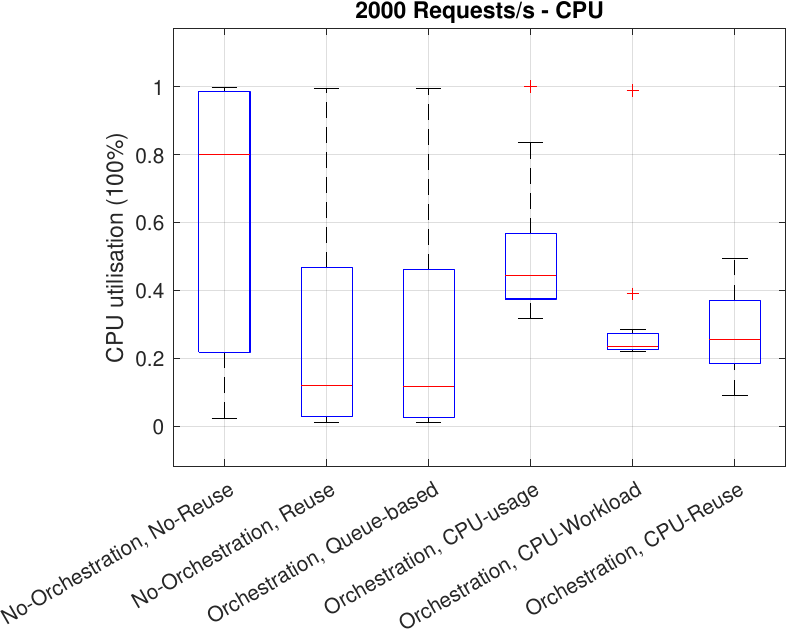}
\centering
\captionsetup{justification=centering}
\centering
\label{subfig:200_CPU_Avg_Box_effective_Traffic}
}
~
\subfigure{
\includegraphics[width=0.3\textwidth]{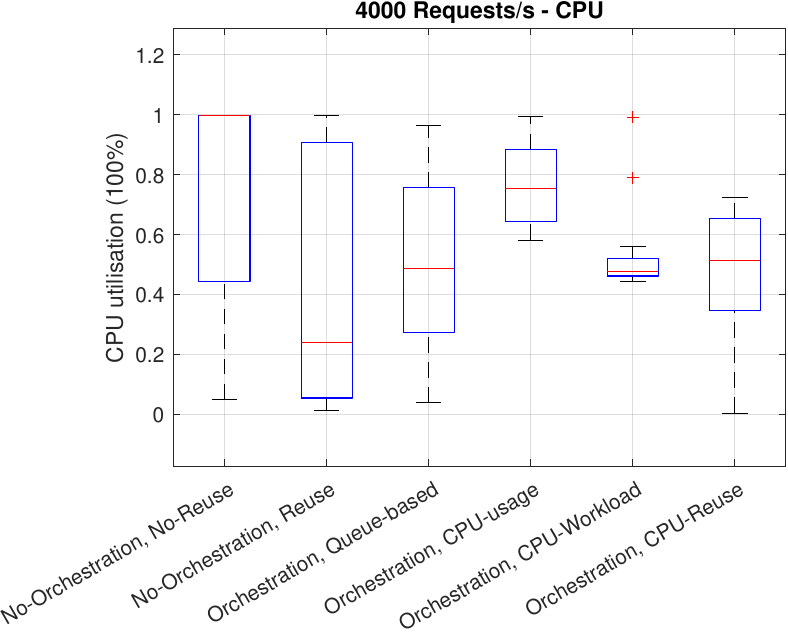}
\centering
\captionsetup{justification=centering}
\label{subfig:CPU4000_Traffic}
} 
~
\subfigure{
\includegraphics[width=0.3\textwidth]{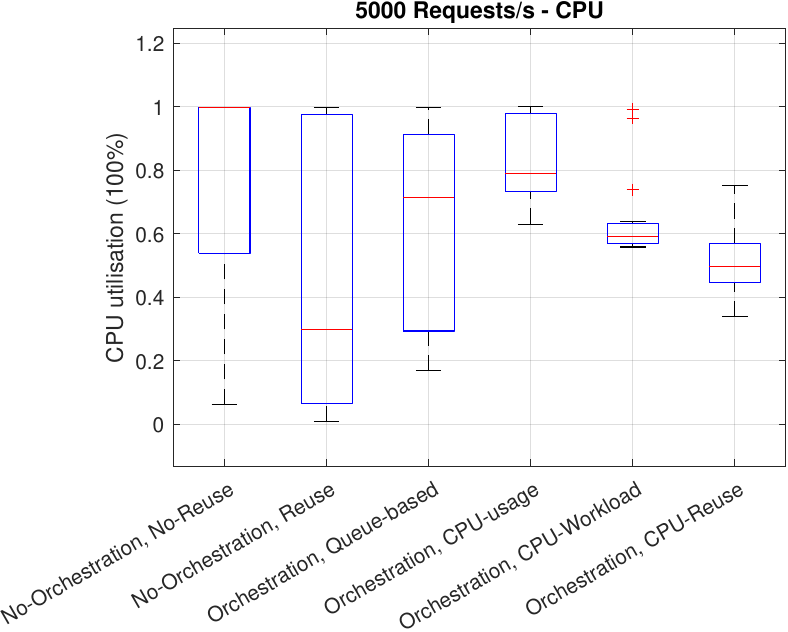}
\centering
\captionsetup{justification=centering}
\label{subfig:5000_CPU_Avg_Box_effective_Traffic}
}
\vspace{-0.2cm}
\caption{CPU utilisation for different rates with Traffic Detection Dataset.}
\label{fig:CPU_Traffic}
\vspace{-0.2cm}
\end{figure*}

\begin{figure*}[h]
\centering
\subfigure{
\includegraphics[width=0.3\textwidth]{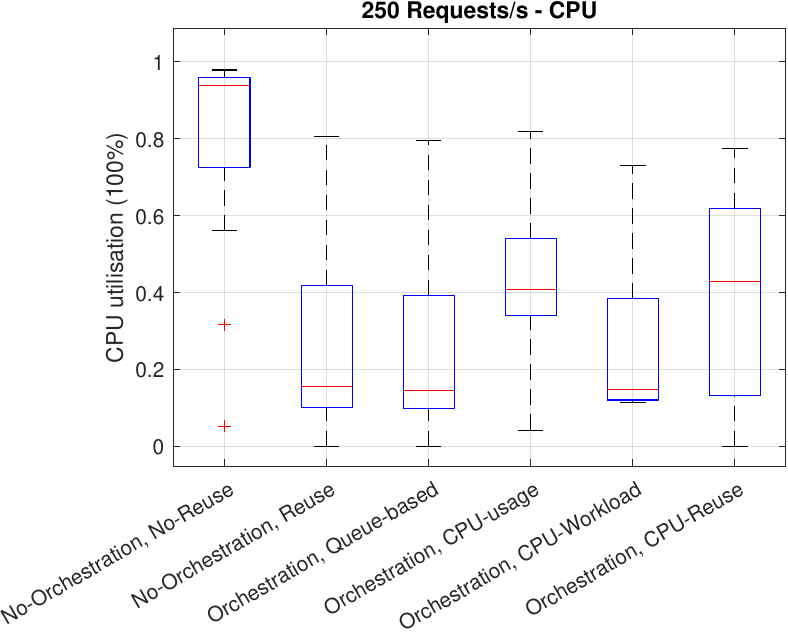}
\centering
\captionsetup{justification=centering}
\centering
\label{subfig:150_CPU_Avg_Box_effective_Alexa}
}
~
\subfigure{
\includegraphics[width=0.3\textwidth]{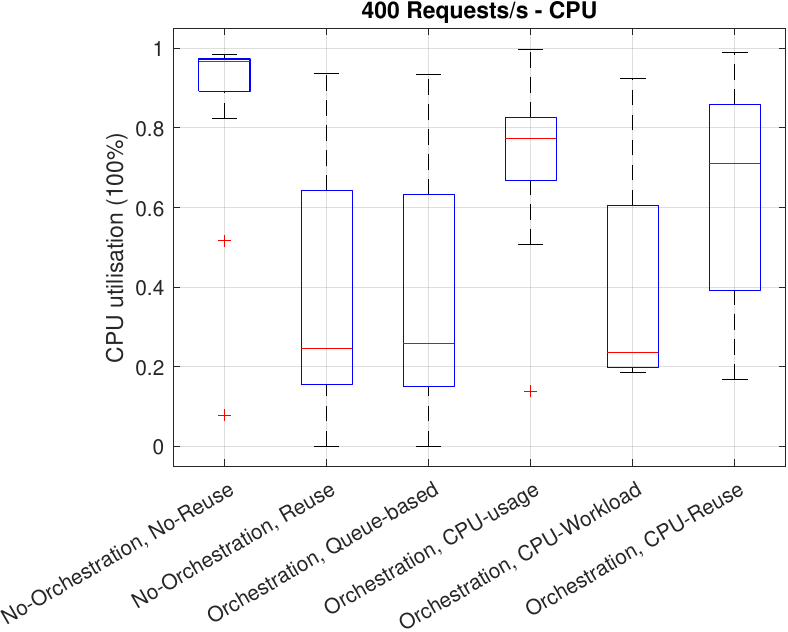}
\centering
\captionsetup{justification=centering}
\label{subfig:300_CPU_Avg_Box_effective_Alexa}
} 
~
\subfigure{
\includegraphics[width=0.3\textwidth]{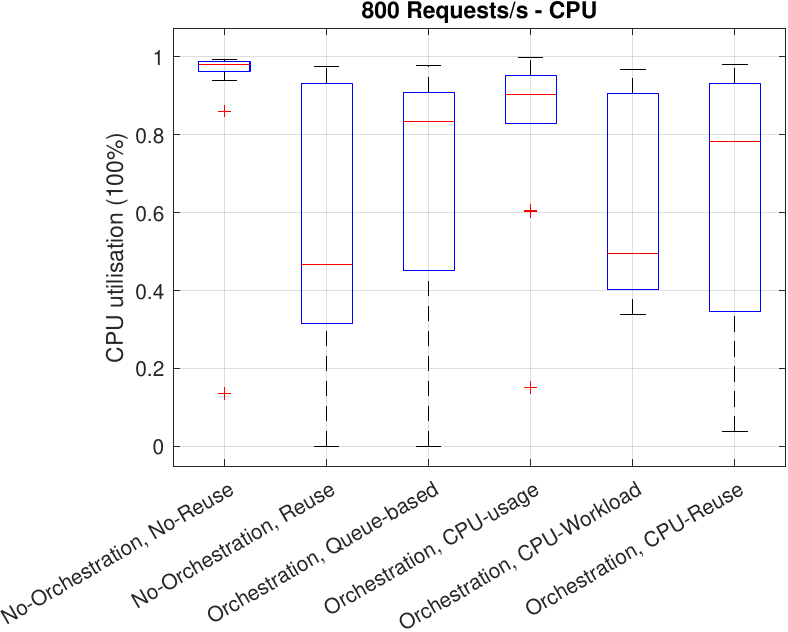}
\centering
\captionsetup{justification=centering}
\label{subfig:500_CPU_Avg_Box_effective_Alexa}
}
\vspace{-0.2cm}
\caption{CPU utilisation for different rates with Alexa Dataset.}
\label{fig:CPU_Alexa}
\end{figure*}

\begin{figure*}[h]
\centering
\subfigure{
\centering
\includegraphics[width=0.3\textwidth]{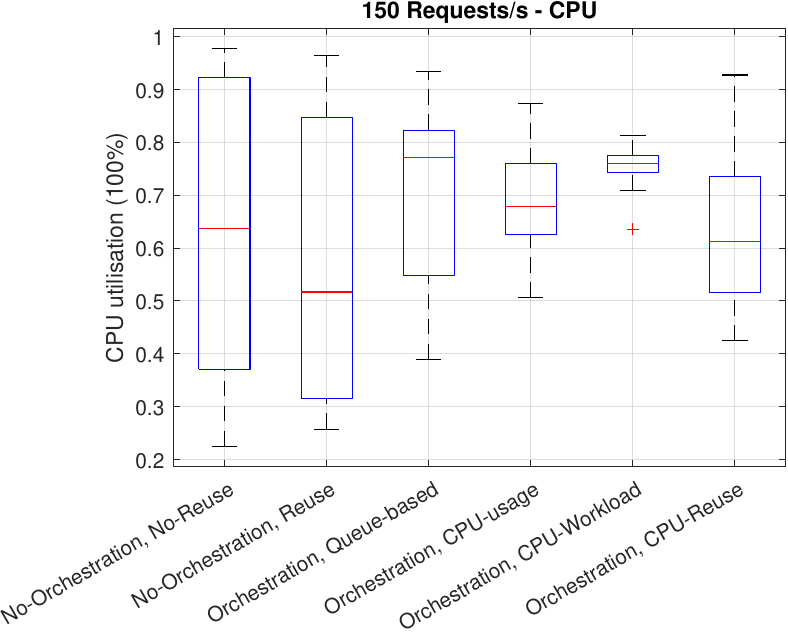}
\centering
\captionsetup{justification=centering}
\centering
\label{subfig:150_CPU_Avg_Box_effective_General}
}
~
\subfigure{
\includegraphics[width=0.3\textwidth]{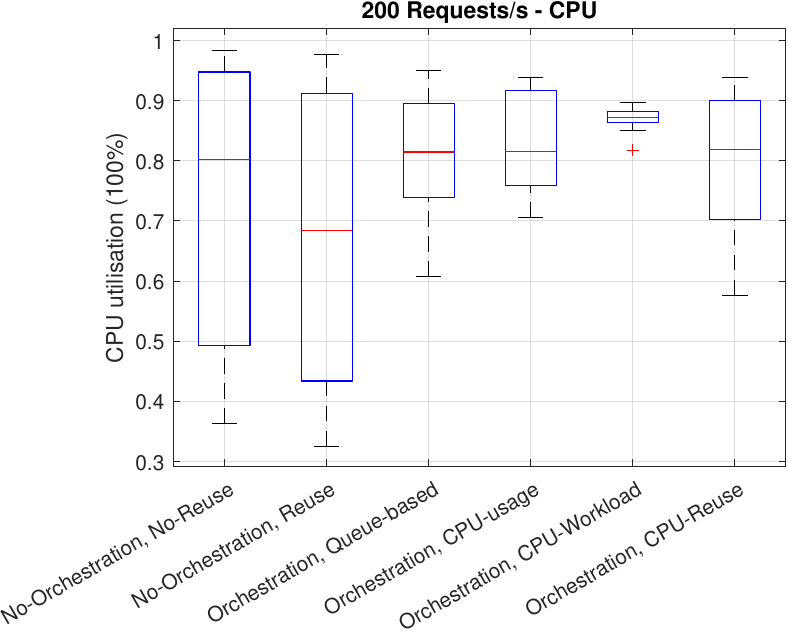}
\centering
\captionsetup{justification=centering}
\label{subfig:200_CPU_Avg_Box_effective_General}
} 
~
\subfigure{
\includegraphics[width=0.3\textwidth]{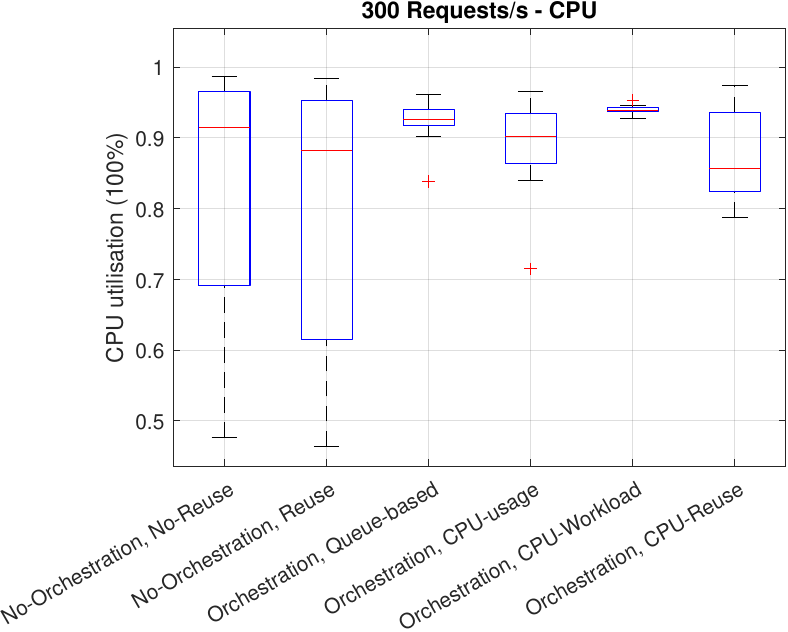}
\centering
\captionsetup{justification=centering}
\label{subfig:300_CPU_Avg_Box_effective_General}
}
\vspace{-0.2cm}
\caption{CPU utilisation for different rates with General Commands Dataset.}
\label{fig:CPU_General}
\vspace{-0.3cm}
\end{figure*}

\emph{CPU utilisation} - Since we started by considering throughput, we shall now look at CPU utilisation, since it is the next resource in line to create bottlenecks in the system. Our main concerns, in the end, are system resource utilisation and transmission performance. In this regard, we saw from the first part that the CPU-usage strategy of MNIST fares very well, even though the Workload strategy came in to a close second. This was due mainly to the low complexity and high diversity of the data.
With the Traffic detection dataset, the situation is wholly different. The CPU-usage is the most balanced and well distributed in the case of the CPU-Reuse strategy, with the CPU-Workload strategy achieving very good load balancing across EDRs (with higher median load in some cases), due to the high reusability rates of the dataset. Note that, while the data is very diverse in setting and quite complex, it also is large in both the number of data points/images, which is why the CPU utilisation is high even in the high-reuse cases.
Thus, the most stable and efficient strategy in the case of the Traffic detection dataset is the CPU-Reuse strategy. In the case of the Alexa dataset, there is a distinct advantage of the CPU-Workload strategy in regards to CPU utilisation, due mainly to the better distribution of CPU-bound requests across EDRs. 
However, the queue-based strategy shows more localised CPU utilisation management (Figure~\ref{fig:CPU_Alexa}), which puts it in a good place, as a contender for the best strategy to work towards the best performance improvement associated with the Alexa dataset, because of less load balancing and more overall CPU-utilisation reduction. This enables some nodes to possibly be engaged (through slicing or simple instantiation within the same domain) into other networked orchestration and/or data processing tasks.
The general commands dataset is where the queue-based orchestration strategy shines brightest, as it provides the best results for this dataset (Figure \ref{fig:CPU_General}). This is due to the high engagement of network interactions, together with CPU utilisation needed for the right distribution of reusable (simplistic/"popular" processed) and processing-bound (previously unprocessed/historically diverse) commands, towards service processing queue optimisation.

\subsection{Evaluation Conclusions and Deployment Options}
We can now draw the main conclusions of the above evaluation, also associating the orchestration strategies with data specific to the different datasets and their associated use cases/user environments, according to the performance improvement exhibited by each strategy.
\begin{enumerate}[leftmargin=0cm,itemindent=.5cm,labelwidth=\itemindent,labelsep=0cm,align=left, noitemsep, topsep=0pt]
    \item The algorithms implemented in this evaluation were four relatively simple strategies that were used to demonstrate the benefits of the idea of orchestration in this edge computation with storage and reuse scenario, with different environments and use cases in mind, and their refinement to improve performance further is a subject for future work and engineering.
    \item The four orchestration strategies showed improvements over the simple reuse case in terms of CPU load balancing and request throughput. The degree of improvement of each strategy varied with the application use case due to the computational complexity of processing queries for that application, the nature of the served data and the degree of query similarity in that use case which impacts the amount of reuse that is possible. A single orchestration strategy did not perform best in all combinations of use case and workload level, which indicates that a combination of orchestration strategies would be beneficial and that designers of future algorithms should consider multiple factors:
    \begin{itemize}[leftmargin=0cm,itemindent=.3cm,labelwidth=\itemindent,labelsep=0cm,align=left, noitemsep, topsep=0pt]
        \item Resource Utilisation - this represents and is represented by both processing resources and storage. We only analyse processing resources in this case, as storage is less affected by the cases studied here, due to reuse;
        \item User-/Application Provider-imposed QoE and QoS - these service metrics could be integrated as "tuning knobs", in potential designs and implementations of the system;
        \item Orchestration Overhead - all the management and optimisation comes at a price, and the price is comprised of more computing resource utilisation, combined with timing constraints and a small, periodic routing overhead. 
    \end{itemize}
\end{enumerate}

Looking at the system, as a whole, and with the above in mind, we conclude that, while reuse reduces the processing load on EDRs and hence increases the throughput of requests, orchestration improves the distribution of load amongst the EDRs and further increases request throughput which improves the QoE of the users and better manages system resources, for more resource utilisation control. However, as a secondary overall evaluation conclusion, we would like to note that equal load distribution within the whole system may not be the most desirable outcome, considering service and/or data and/or privacy and/or security needs heterogeneity, within each system. 

One option for deployment of the orchestration algorithms presented in this paper is to implement them as part of the Network Function Virtualisation (NFV) architecture~\cite{VNF} as an application in the Management and Orchestration (MANO) framework~\cite{OSM}. A MANO environment of kubernetes clusters or OpenStack-managed servers would provide a suitable environment for the execution of optimisation functions that: 1) gather usage and performance data from distributed EDRs; 2) calculate bucket placement strategies based on the collected operational data; and 3) implement the movement of buckets and caches between EDRs and to configure the associated routing table changes.

\section{Conclusion and Future Work}
\label{sec:conclusion}
This paper presents \sol: an LSH- and Storage-based Computation and Reuse management framework. This framework's main purpose is to improve on all resource utilisation and QoS performance of the system, by exploiting its addressing, storage, computing and data reusability capabilities and potential. Its realisation is possible through the deployment of EDRs, starting at 1-hop away from users, seamlessly interacting with network traffic originating at the edge. Through a logically centralised management plane, \sol implements orchestration strategies, towards better function and data placement, based on LSH-obtained data hashes and their associated buckets. For the system's evaluation, we implemented a local EDR, with LSH, similarity checks and storage, for the purpose of demonstrating the mechanics for each dataset. We show that the orchestrated system is efficient, providing a realistic scenario for a large scale, simulated environment, with EDRs co-located with gateways, to prove the feasibility and analyse the performance of the developed orchestration strategies.

In the future, we plan to explore the following directions: (i) the impact on energy efficiency and sustainability; (ii) mechanisms to enable the collaboration among EDR providers; (iii) mechanisms to enable capitalisation of reuse, in conjunction with storage and their orchestration; iv) the integration of intelligence, towards the detection of patterns, (re)association and (re)definition of optimisation algorithms, for cross-application classes and (v) the deployment of \sol in large-scale settings, where it can interact and enable the operation of real-world applications.

\bibliographystyle{IEEEtran}
\bibliography{sample-bibliography}

\vspace{-0.8cm}
\begin{IEEEbiographynophoto}{Adrian-Cristian (Chris) Nicolaescu} has recently received his PhD in Electronics and Electrical Engineering, concentrating on "Computing at the Edge of the Internet", from University College London (UCL) and a Senior Research Associate, within the High Performance Networks (HPN) Research Group of the Smart Internet Lab, at the University of Bristol. His research interests mainly include edge computing, decentralised and distributed systems and edge-based storage systems.
\end{IEEEbiographynophoto}

\vspace{-0.8cm}
\begin{IEEEbiographynophoto}{Spyridon Mastorakis} is an Assistant Professor in Computer Science and Engineering, at the University of Notre Dame and has previously served as an Assistant Professor in the Computer Science Department of the University of Nebraska at Omaha, where he was the Director of the Ph.D. program in Information Technology, Director of the Network Systems Research laboratory, and a co-Director of the Robotics, Networking, and Artificial Intelligence laboratory.
\end{IEEEbiographynophoto}

\vspace{-0.8cm}
\begin{IEEEbiographynophoto}{Md Washik Al Azad} is currently pursuing a Ph.D. degree in Information Technology at the University of Notre Dame. He received his Bachelor's degree in Electronics \& Telecommunication Engineering (ETE) from Rajshahi University of Engineering \& Technology, Bangladesh. His research interests include edge computing, network systems, and security. Currently, He is working at the Network Systems Research laboratory at UND, under the supervision of Dr. Spyridon Mastorakis.
\end{IEEEbiographynophoto}

\vspace{-0.8cm}
\begin{IEEEbiographynophoto}{David Griffin} is a Principal Research Associate in the Communications and Information Systems Group, Department of Electronic and Electrical Engineering, University College London. He has a BSc in Electronic, Computer and Systems Engineering from Loughborough University and a PhD in Electronic and Electrical Engineering from UCL
\end{IEEEbiographynophoto}

\vspace{-0.8cm}
\begin{IEEEbiographynophoto}{Miguel Rio} is a Professor of Computer Networks in the Department of Electronic and Electrical Engineering at University College London. He holds a Meng and MSc from the University of Minho, Portugal and a Phd from the University of Kent, United Kingdom. He has been a principle investigator in numerous research projects funded by the UK government, the European Union, industry and the US/UK military. His current interests are on edge networking, software defined networking, network resilience and improving network quality of experience.
\end{IEEEbiographynophoto}

\end{document}